\documentclass{ws-mpla}
\usepackage[super]{cite}
\usepackage{graphicx}
\begin{document}


\catchline{}{}{}{}{}

\title{On the Time Dependent  Rindler Hamiltonian Single Particle Eigen States in Momentum Space}

\author{\footnotesize Soma Mitra}

\address{$^a$Department of Physics, Visva-Bharati, Santiniketan, India 731235\\
E-mail: somaphysics@gmail.com
}
\author{\footnotesize Sanchita Das}

\address{$^a$Department of Physics, Visva-Bharati, Santiniketan, India 731235\\
E-mail: sanchedasm@gmail.com
}
\author{Somenath Chakrabarty$^a$}

\address{$^a$Department of Physics, Visva-Bharati, Santiniketan, India 731235\\
E-mail: somenath.chakrabarty@visva-bharati.ac.in
}

\maketitle

\pub{Received (Day Month Year)}{Revised (Day Month Year)}
\begin{abstract}
We have developed a formalism to obtain
the time evolution of the eigen states of Rindler Hamiltonian in momentum space. We have
discussed the difficulties with characteristic curves, and re-cast the time evolution equations in the form of
two-dimensional Laplace equation. The solutions are obtain both in
polar coordinates as well as  in the Cartesian form. It has been
observed that in the Cartesian coordinate, the probability density is
zero both at $t=0$ (the initial time) and at $t=\infty$ (the final time)
for a give $x$-coordinate. The reason behind such peculiar behavior of
eigen state is because it satisfies $1+1$-dimensional Laplace equation.
\end{abstract}

\ccode{PACS Nos.: 
03.65.Ge,03.65.Pm,03.30.+p,04.20.-q 
}
\section{Introduction} Exactly like the flat Minkowski space-time geometry, the Rindler
space is also locally flat \cite{R1,R2,R3,R4,R4a,R5,R6,R6a}. 
The only
difference is that the observer is moving with uniform acceleration
\cite{R1,R2,R3,R4,R4a,R5,R6,R6a}.
Then according to the principle of equivalence,
the Rindler space corresponds to a frame at rest in presence of a locally uniform gravitational field. The strength
of the gravitational field is exactly equal to the magnitude of the
local acceleration \cite{R1,R2,R7,R7a}. The
space-time transformations in Rindler space are given by
\begin{eqnarray}
ct&=&\left (\frac{c^2}{\alpha}+x^\prime\right )\sinh\left (\frac{\alpha t^\prime}
{c}\right ) ~~{\rm{and}}~~ \nonumber \\
x&=&\left (\frac{c^2}{\alpha}+x^\prime\right )\cosh\left (\frac{\alpha t^\prime}
{c}\right ) \nonumber
\end{eqnarray}
Hence the Rindler metric can be written as
\[
g^{\mu \nu} \equiv {\rm{diag}}\left( \left ( 1 + \frac{\alpha x}{c^2} 
\right ) , -1, -1,-1  \right)
\]
In $1+1$ dimension it reduces to
\[
g^{\mu \nu} \equiv {\rm{diag}}\left( \left ( 1 + \frac{\alpha x}{c^2} 
\right ) , -1  \right)
\]
Then with some simple algebraic manipulation it can very easily be shown that 
the single particle Hamiltonian in Rindler space in
one-dimensional form is given by \cite{R4a,R5,R6,R6a}
\begin{equation}
H=\left ( 1+\frac{\alpha x}{c^2}\right )(p^2c^2+m_0^2c^4)^{1/2}
\end{equation}
where $\alpha$ is the uniform acceleration of the frame or
the uniform gravitational field in the neighborhood of the point indicated
by the coordinate $x$. Here $p$ and $m_0$ are the particle momentum along the positive $x$-direction 
and the rest mass respectively and $c$ is the speed
of light in free space.

Further, the observer is assumed to have one dimensional motion along positive $x$-direction. Then in the
non-relativistic approximation, neglecting the rest mass energy term, the Hamiltonian given by eqn.(1) reduces to
\begin{equation}
H=\left ( 1+\frac{\alpha x}{c^2}\right )\frac{p^2}{2m_0}
\end{equation}
In this article we shall investigatethe time evolution of wave function un the momentum space and finally taking the
Fourier transform of the momentum dependent we have obtained the actual wave function as the function of space-time
coordinates.We have organized the paper in the following manner: In the next section, we have developed the basic
formalism. In section 3, we have obtained the wavefunction in space-time coordinate ($x,t$) and finally in section 4, we
have discussed the conclusion of this work.

\section{Basic Formalism}
Now in the quantum mechanical scenario, the classical dynamical variables $x$, $p$ and $H$ are treated as
operators and the canonical quantization condition $[x,p]=i\hbar$ is
satisfied. The aim of this article is to study the evolution of Rindler
Hamiltonian eigenstate with time in momentum space in $1+1$ dimension.
This is the continuation of some of our previously reported results
\cite{R5,R6,SMSC}.
Then following the article by R.W.
Robinett in Am. Jour. Phys. \cite{RO} (see also \cite{RO1,RO2}) the
time dependent Schr\"odinger equation with the momentum representation of the positional coordinate operator $x$
($x=i\hbar \partial/\partial p$) is given by
\begin{equation}
\left ( 1+\frac{\alpha}{c^2}i\hbar\frac{\partial}{\partial p}\right ) \frac{p^2}{2m_0}\psi(p,t)=i\hbar
\frac{\partial\psi(p,t)}{\partial t}
\end{equation}
After the substitution of
\begin{equation}
\frac{p^2}{2m_0}\psi(p,t)=\phi(p,t)
\end{equation}
it becomes
\begin{equation}
\phi(p,t)+i\hbar \frac{\alpha}{c^2}\frac{\partial \phi(p,t)}{\partial p}=i\hbar \frac{2m_0}{p^2} \frac{\partial
\phi(p,t)}{\partial t}
\end{equation}
To solve this equation analytically, we multiply both sides of the above differential equation by the integrating factor
\begin{equation}
I_p=\exp\left (\frac{i\hbar \alpha}{c^2}p\right )
\end{equation}
Then the above Schr\"odinger equation reduces to
\begin{equation}
\frac{\partial \xi}{\partial \tau}-i\frac{\partial \xi}{\partial q}=0
\end{equation}
where 
\begin{equation}
\xi(q,\tau)=I_p\phi(p,t), ~~\tau=\frac{t}{2m_0\hbar} ~~{\rm{and}}~~ q=\frac{1}{p}
\end{equation}
Let us now try to solve this reduced form of the Schr\"odinger equation given by eqn.(7). We first try with the well
known characteristics flow equation approach \cite{R8}. We introduce two new variables
\begin{equation}
r=\tau +iq ~~ {\rm{and}}~~ s=-i\tau -q
\end{equation}
Then the above differential equation reduces to
\begin{equation}
\frac{\partial \xi(r,s)}{\partial s}=0
\end{equation}
which gives $\xi(\tau,q)=f(r)=f(\tau+iq)$.
Here $f(\tau+iq)$ is completely arbitrary. 
The arbitrariness of the solution can sometime be removed by the
boundary conditions. However, the complex characteristics given by
eqn.(10) has no practical value. Also it has no obvious relevance to
boundary conditions. Moreover, the propagation along such complex
characteristics will not provide any acceptable partial differential
equation for a real physical system. Or in other wards there will be no acceptable form of differential equation to
represent the flow with physical boundary conditions. Therefore we shall try an
alternative approach to get the solution. 

We have now followed the technique to obtain Cauchy-Euler differential
equation \cite{R8,R9}.
We take derivatives of both the terms of eqn.(7) with respect to
$\tau$ and $iq$ and then adding these two equations, we get
\begin{equation}
\frac{\partial^2\xi}{\partial \tau^2}+ \frac{\partial^2 \xi}
{\partial q^2}=0
\end{equation}
This is the well known Laplace equation in two-dimension. We may call the above equation  as the circular type
partial differential equation. To get solutions of this equation
we consider  circular polar coordinate $(\rho, \phi)$, where
$\tau=\rho \cos\phi$, $q=\rho\sin\phi$ and $\phi=\tan^{-1}(q/\tau)=\tan^{-1}({\rm{constant}}/pt)$, where the
constant$=2m_0\hbar$. Then
the above
differential equation reduces to
\begin{equation}
\frac{1}{\rho}\frac{\partial}{\partial \rho}
\left (\rho \frac{\partial \xi}{\partial \rho}\right
)+
\frac{1}{\rho^2} \frac{\partial^2 \xi}{\partial\phi^2} =0
\end{equation} 
With the separable form of wave function, given by $\xi(\rho,\phi)
=R(\rho)\Phi(\phi)$, we have
\begin{equation}
\frac{d^2\Phi}{d\phi^2}+\nu^2 \Phi=0
\end{equation}
and
\begin{equation}
\rho\frac{d}{d\rho}\left (\frac{d R}{d\rho}\right )-\nu^2 R=0
\end{equation}
where $\nu$ is an unknown real constant.
Here the first one is the well known differential equation for
classical harmonic oscillator in one dimension, whereas the second
one is called the Cauchy-Euler equation. The solution of the first
equation is
\begin{equation}
\Phi(\phi)=A\cos(\nu \phi)+B \sin(\nu \phi)
\end{equation}
with $A$ and $B$ are two unknown constants. The solutions of the
Cauchy-Euler equation are known to be of the form $\rho^\gamma$.
Substituting in eqn.(14), we have $\gamma=\pm\nu$.
Hence for the radial part we may write
\begin{equation}
R(\rho)=a\rho^\nu+b\rho^{-\nu}
\end{equation}
where $a$ and $b$ are two other unknown constants. 
In fig.(1) we have plotted for the sake of illustration the wave function $\xi(\rho,\phi)$ for all the constants
equal to unity along with $\nu=1$. The variable $\phi$ is along $x$-axis and $\rho$ is along $y-axis$. One should
note that from the definitions of $\phi$ and $\rho$ it is quite obvious that both of them are positive in nature.
\begin{figure}[ph]
\centerline{\includegraphics[width=2.0in,angle=-90]{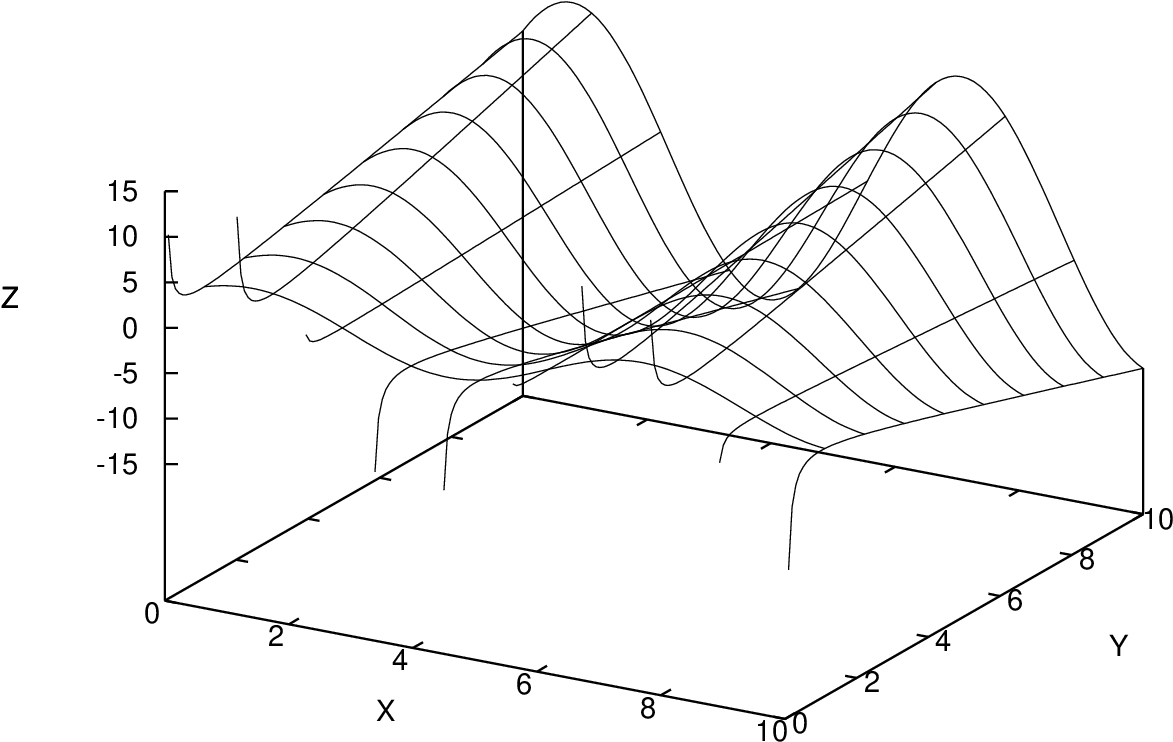}}
\vspace*{8pt}
\caption{Variation of $\xi(\rho,\phi)$ with $\phi$, along $x$-axis and $\rho$ along $y$-axis.
\protect\label{fig1}}
\end{figure}

As a special case, for
$\nu=0$
\begin{equation}
\Phi(\phi)=A_0+B_0\phi
\end{equation}
and
\begin{equation}
R(\rho)=a_0+b_0\ln \rho
\end{equation}
In fig.(2) we have plotted the wave function $\xi(\rho,\phi)$ for this special case.
\begin{figure}[ph]
\centerline{\includegraphics[width=2.0in,angle=-90]{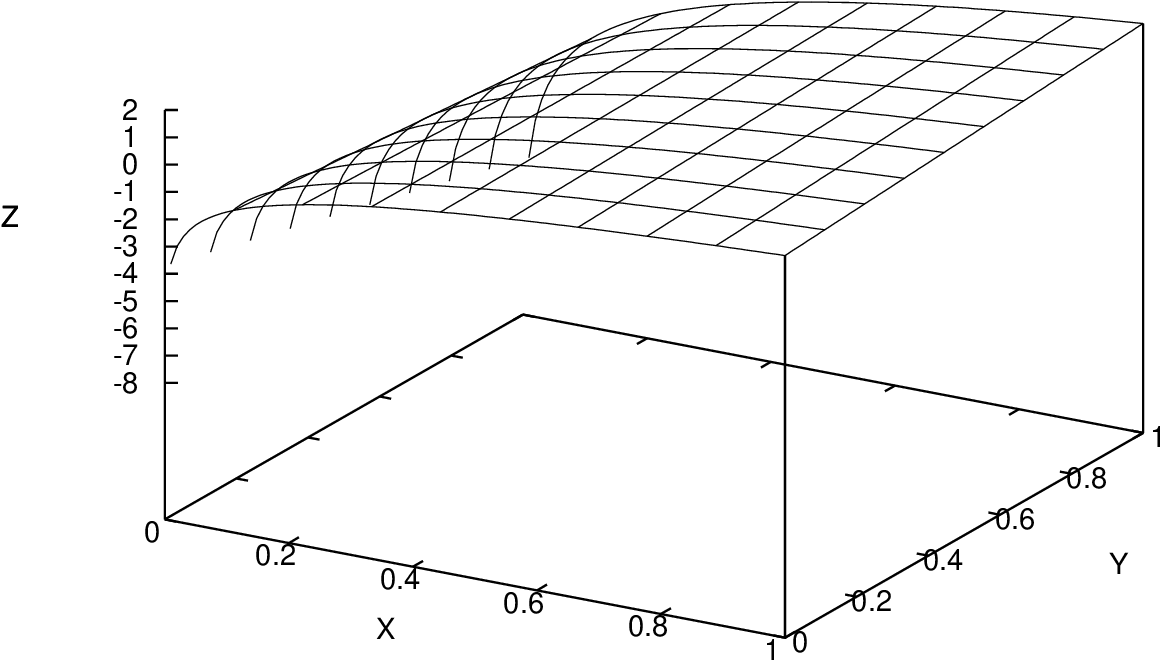}}
\vspace*{8pt}
\caption{The surface plot of the wave function. $\phi$ is along $x$-axis and $\rho$ along $y$-axis.
\protect\label{fig2}}
\end{figure}
In the electrostatic problem the above solutions are generally considered as the building blocks  to construct the
potential by linear superposition. The radial logarithmic form gives the potential due to a linear charge
distribution. Taking into account this concept of superposition principle, we can
also write down the general solution for the wave functions. However, in the present situation, $\phi$ is
restricted in the first quadrant only ($0\leq \phi \leq \pi/2$).
Therefore, $\nu$ does not necessarily have integer values.
Therefore no further progress can be achieved from this approach.

Next we shall try to get solutions in $\tau-q$ space, which is equivalent to two dimensional ($x-y$) Cartesian
coordinate system. Using the separable form of solutions $\xi(\tau,q)=T(\tau)Q(q)$, we have
\begin{equation}
\frac{d^2T}{d\tau^2}+\omega^2T=0
\end{equation}
and
\begin{equation}
\frac{d^2Q}{dq^2}-\omega^2 Q=0
\end{equation}
where $\omega$ is an unknown real constant. It is trivial to show that the solutions are
\begin{equation}
T(\tau)=C_1\sin \omega \tau+ C_2\cos \omega \tau
\end{equation}
and
\begin{equation}
Q(q)=D_1\sinh\omega q+D_2\cosh\omega q
\end{equation}
Where $C_1$, $C_2$, $D_1$ and $D_2$ are unknown constants. Since $\tau$ is proportional to the actual time
coordinate $t$, which increases continuously, and not periodic or bounded, there is no problem in having same
wave function at various time coordinate. Therefore, $\omega$ is not necessarily an integer.  
Whereas $q\propto 1/p$, therefore for large $p$,
$q\longrightarrow 0$ and then neglecting the higher degree terms, $Q\propto q\propto 1/p$. In the case of
$p\longrightarrow 0$, i.e., $q\longrightarrow \infty$, the asymptotic form of the solution $Q\longrightarrow
\infty$, which is unphysical. To make the solution physically
acceptable, we put by hand $D_2=-D_1=-D$, then
\begin{equation}
Q(q)=D(\cosh\omega q-\sinh\omega q)
\end{equation}
Which becomes after re-defining the constant $D$,
\begin{equation}
Q(q)=D\exp(-\omega q)
\end{equation}
This is the asymptotic form of the solution.
Therefore the final solutions for all possible values of 
$\omega$ are given by
$$
\xi(\tau,q)=\sum_\omega (C_1\sin \omega\tau+C_2 \cos \omega \tau)(D_1\sinh \omega q+D_2\cosh \omega q) \eqno(25a)
$$
and asymptotically
$$
\xi(\tau,q)=\sum_\omega (C_1\sin \omega\tau+C_2 \cos \omega \tau)\exp(-\omega q) \eqno(25b)
$$
Assuming that the variable $\omega$ is changing continuously, we can transform the sum over $\omega$ by integral,
and then we can write down the asymptotic form of the wave function after evaluating the integral over $\omega$ 
within the limit $0$ to $\infty$ and is given by:
\begin{equation}
\xi(\tau,q)=\frac{C_1\tau+C_2q}{\tau^2+q^2}
\end{equation}
where $C_1$ and $C_2$ are two re-defined constants to make the expression dimensionally correct.
In fig.(3) we have plotted $\xi(\tau,q)$ for $\omega=1$ and as before for the sake of illustration, all the
constants are also set equal to unity. In fig.(4) we have
shown the same wave function, but in the asymptotic region. Whereas in fig.(5) we have plotted the asymptotic form of
wave function for all possible values of $\omega$, which is assumed to be  changing continuously.
\begin{figure}[ph]
\centerline{\includegraphics[width=2.0in,angle=-90]{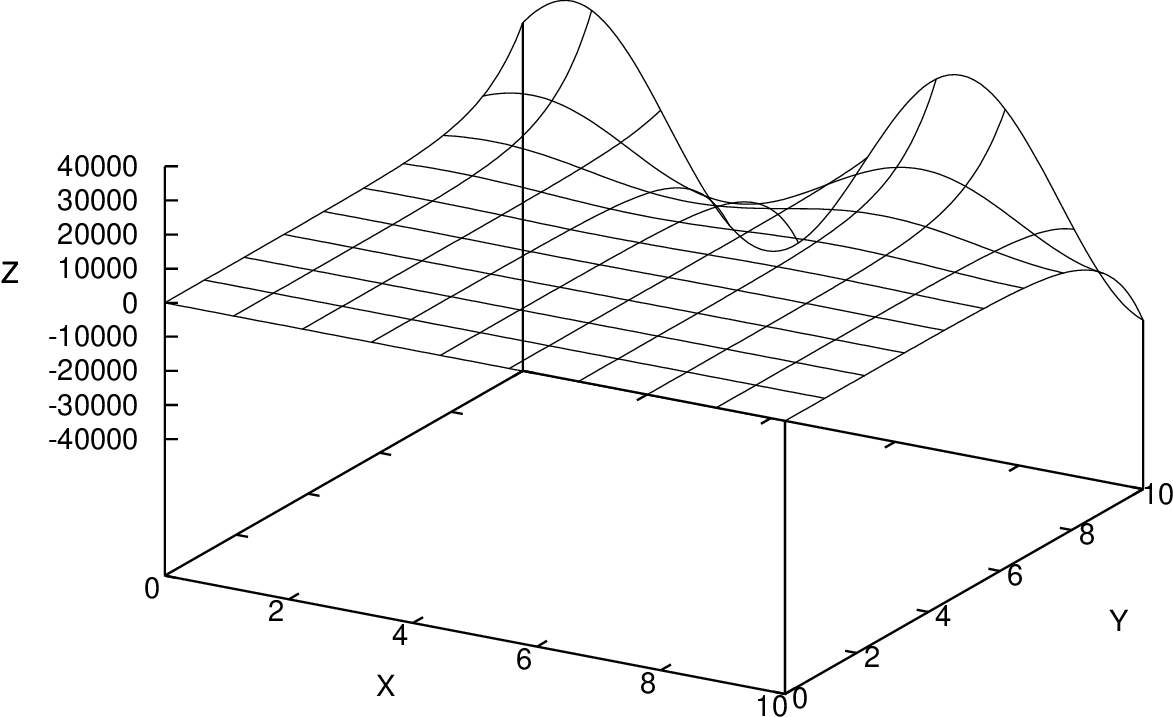}}
\vspace*{8pt}
\caption{The surface plot of the wave function $\xi(\tau,q)$, $tau$ is along $x$-direction and $q$ along $y$-axis.
\protect\label{fig3}}
\end{figure}
\begin{figure}[ph]
\centerline{\includegraphics[width=2.0in,angle=-90]{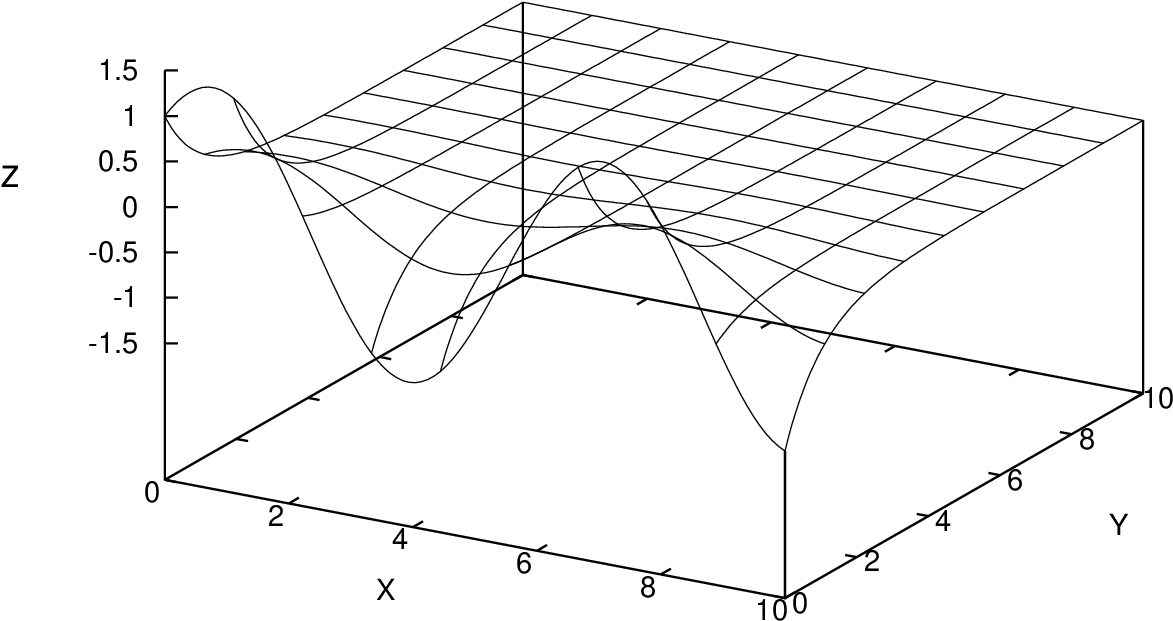}}
\vspace*{8pt}
\caption{The surface plot of the asymptotic form of the wave function  
$\xi(\tau,q)$, $tau$ is along $x$-direction and $q$ along $y$-axis.
\protect\label{fig4}}
\end{figure}
\begin{figure}[ph]
\centerline{\includegraphics[width=2.0in,angle=-90]{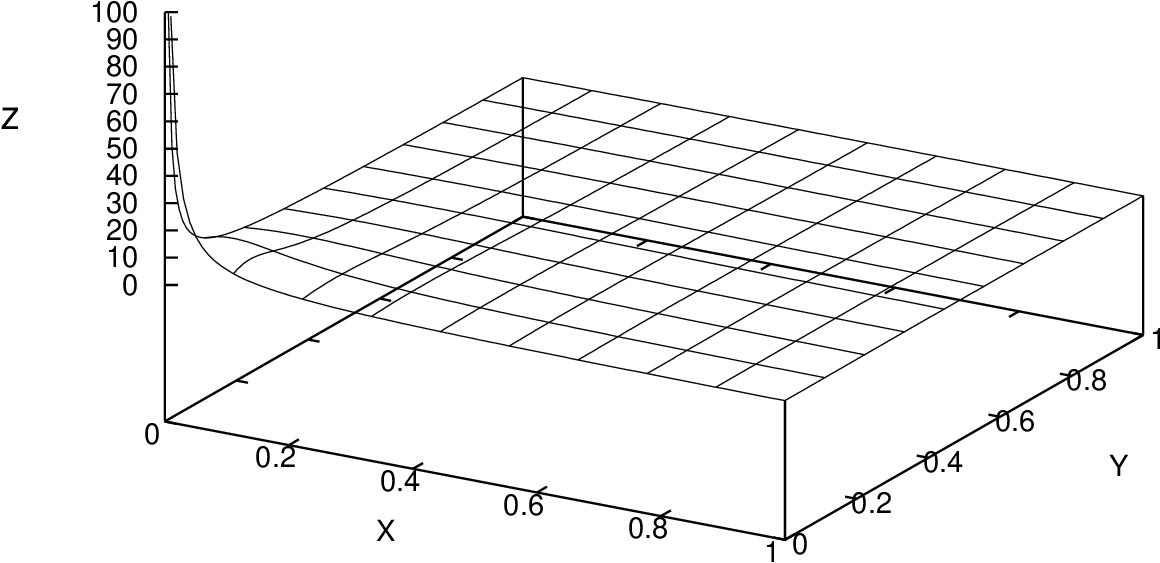}}
\vspace*{8pt}
\caption{The surface plot of the asymptotic form of the wave function for all possible values of $\omega$,  
$\xi(\tau,q)$, $tau$ is along $x$-direction and $q$ along $y$-axis.
\protect\label{fig5}}
\end{figure}

\section{Wavefunctions in Space-Time Coordinates}
Therefore, it is quite obvious that the time dependence of eigen states of the Rindler Hamiltonian is 
quite different from the stationary state solutions of the wave functions.
To make it more clear we consider the Fourier transform of eqn.(25)
to obtain the space-time dependent form of the wave function. The
Fourier transform is given by
\begin{equation}
\psi(x,t)=\frac{N}{2\pi}\int_{-\infty}^{+\infty}
\frac{at+\frac{b}{p}} {a^2t^2+\frac{b^2}{p^2} } \exp(-ipx)dp
\end{equation}
where $N$ is the normalization constant.
On evaluating the Fourier transform, we get 
\begin{equation}
\psi(x,t)=N\frac{b}{2^{1/2}a^2t^2} \exp\left
(-\frac{bx}{at}-i\frac{\pi}{4} \right )
\end{equation}
The normalized form of the wave function is given by
\begin{equation}
\psi(x,t)=\left(\frac{2b}{at}\right)^{1/2}  \exp\left
(-\frac{bx}{at}-i\frac{\pi}{4} \right )
\end{equation}
The probability density is given by
\begin{equation}
\mid \psi(x,t)\mid^{1/2} =\frac{2b}{at} \exp\left
(-\frac{2bx}{at}\right )
\end{equation}
Obviously the probability density, i.e., the probability of existence 
of the
particle becomes zero for both $t\longrightarrow 0$ and $t\longrightarrow
\infty$. Then in between, there is some specific value of time at which the
probability of existence becomes maximum, given by 
\begin{equation}
\frac{d}{dt}\mid \psi \mid^2=0
\end{equation}
This condition gives 
\begin{equation}
t=\frac{2bx}{a}=t_{\rm{max}}
\end{equation}
This is the time at which the probability of existence becomes
maximum. In fig.(6) we have plotted the scaled quantity
\begin{equation}
\frac{\mid\psi(x,t)\mid}{(2b/a)^{1/2}} =\frac{ \exp(-\gamma/t)}
{t^{1/2}}
\end{equation}
where $\gamma=bx/a$. In fig.(6) top one curve is for $\gamma=0.1$,
rest are in sequence $\gamma=0.5$, $1$, $5$, $10$ and $25$
respectively. The type of time dependence as shown above
is because of faact that the wave functions satisfy the Laplace equation (eqn.(11)) in two-dimension.
\begin{figure}[ph]
\centerline{\includegraphics[width=2.0in,angle=-90]{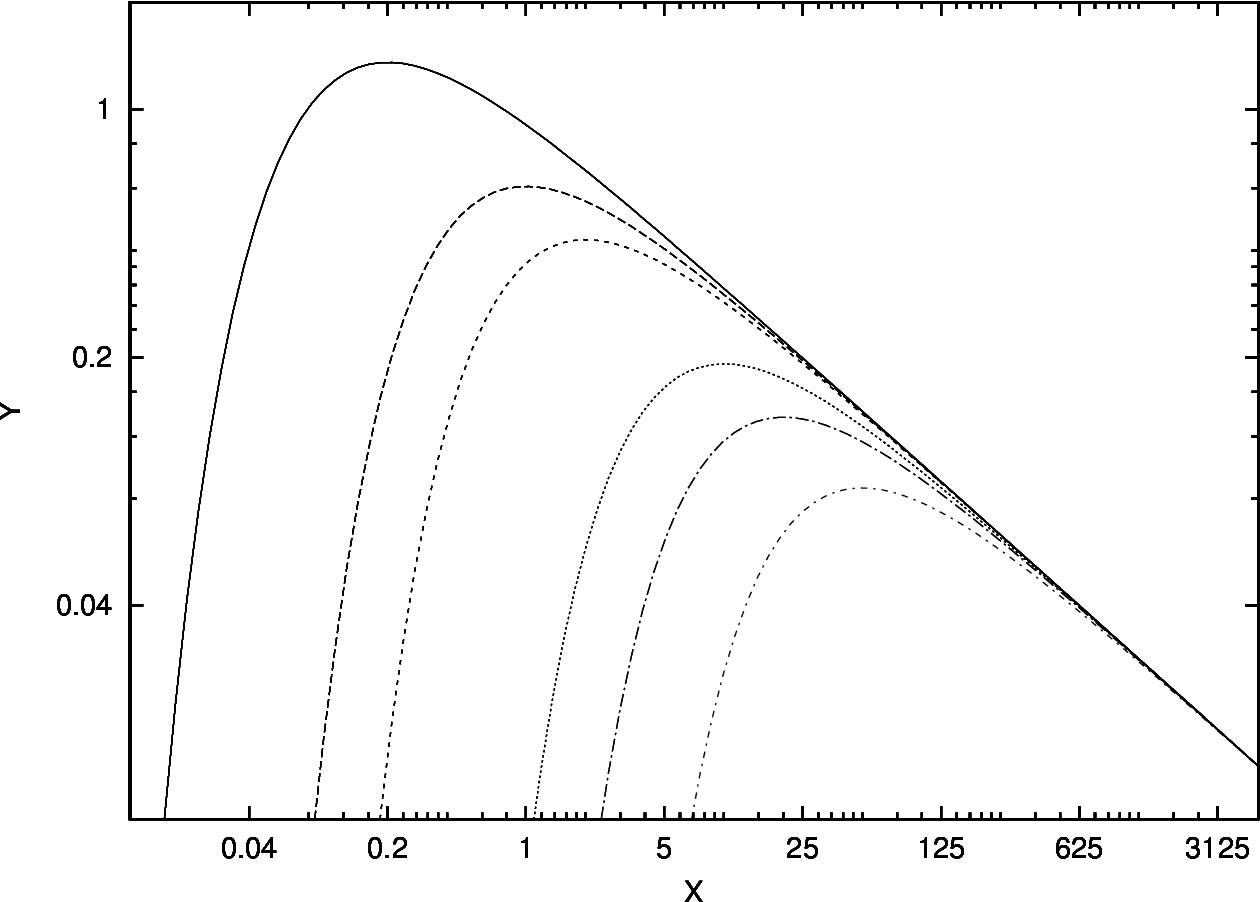}}
\vspace*{8pt}
\caption{The variation of the scaled $\mid \psi(x,t)\mid$ (along
$y$-axis) with time $t$ (along $x$-axis) for various values of
$\gamma$ as mentioned in the text.
\protect\label{fig6}}
\end{figure}
\section{Conclusion}
Finally we conclude by saying that the time dependencies of the Rindler Hamiltonian eigen state is quite different
from the usual stationary state wave functions. 

\end{document}